\documentclass[floats,floatfix,amssymb,prd,twocolumn,superscriptaddress,nofootinbib,preprintnumbers]{revtex4-1}
		
\usepackage{amssymb,amsmath,verbatim,mathtools,needspace,enumitem,etoolbox,graphicx,physics,microtype,afterpage,bm}
\usepackage[dvipsnames, usenames]{xcolor}
\definecolor{linkcolor}{rgb}{0.0,0.3,0.5}
\usepackage[unicode, colorlinks=true, linkcolor=linkcolor, citecolor=linkcolor, filecolor=linkcolor,urlcolor=linkcolor, pdfusetitle]{hyperref}
\usepackage[all]{hypcap}
\usepackage[T1]{fontenc}
\usepackage[utf8]{inputenc}
\usepackage{tabularx}
\usepackage{float}
\usepackage{lmodern}
\allowdisplaybreaks
\usepackage{tikz}
\usepackage{color}
\usepackage{framed}
\usepackage{hyperref}
\hypersetup{colorlinks, citecolor=bluscuro, linkcolor=black, urlcolor=bluscuro}
\definecolor{rossos}{cmyk}{0,1,1,0.55}
\definecolor{bluscuro}{rgb}{0.15, 0.2, .85}
\definecolor{bluchiaro}{cmyk}{1,.3,0.,0.1}
\definecolor{ForestGreen}{rgb}{0.13, 0.55, 0.13}

\newcommand{\be}{\begin{equation}}
\newcommand{\ee}{\end{equation}}

\def\BH{\text{\tiny BH}}
\newcommand{\llp}{\left [}
\newcommand{\rrp}{\right ]}
\newcommand{\lp}{\left (}
\newcommand{\rp}{\right )}

\def\lsim{\mathrel{\rlap{\lower4pt\hbox{\hskip0.5pt$\sim$}}
    \raise1pt\hbox{$<$}}}         %less than or approx. symbol
\def\gsim{\mathrel{\rlap{\lower4pt\hbox{\hskip0.5pt$\sim$}}
    \raise1pt\hbox{$>$}}}         %greater than or approx. symbol

\begin{document}

\title{Modelling frequency-dependent tidal deformability\\
for environmental black-hole mergers
}

\author{Valerio De Luca}
\email{vdeluca@sas.upenn.edu}
\affiliation{Center for Particle Cosmology, Department of Physics and Astronomy, University of Pennsylvania 209 S. 33rd St., Philadelphia, PA 19104, USA}

\author{Andrea Maselli}
\email{andrea.maselli@gssi.it}
\affiliation{Gran Sasso Science Institute (GSSI), I-67100 L'Aquila,  Italy}
\affiliation{INFN, Laboratori Nazionali del Gran Sasso, I-67100 Assergi, Italy}

\author{Paolo Pani}
\email{paolo.pani@uniroma1.it}
\affiliation{Dipartimento di Fisica, Sapienza Università 
	di Roma, Piazzale Aldo Moro 5, 00185, Roma, Italy}
\affiliation{INFN, Sezione di Roma, Piazzale Aldo Moro 2, 00185, Roma, Italy}

% \date{\today}

\begin{abstract}
Motivated by events in which black holes can lose their environment due to tidal interactions in a binary system, we develop a waveform model in which the tidal deformability interpolates between a finite value (dressed black hole) at relatively low frequency and a zero value (naked black hole) at high frequency. We then apply this model to the example case of a black hole dressed with an ultralight scalar field and investigate the detectability of the tidal Love number with the Einstein Telescope. We show that the parameters of the tidal deformability model could be measured with high accuracy, providing a useful tool to understand dynamical environmental effects taking place during the inspiral of a binary system.
\end{abstract}

\preprint{ET-0295A-22}
\maketitle

\section{Introduction}
%----------------------------------------------------------------------------------------------------
Tidal interactions in close astrophysical systems carry golden 
information on the internal composition of compact objects. They 
affect the dynamics of binary sources, leaving a footprint within 
the emitted signals, both in the gravitational-wave~(GW) and in the 
electromagnetic spectrum~\cite{poisson_will_2014}. Pioneering 
calculations, performed at the beginning of the twentieth century, have paved the ground for a rigorous analytical description of tidal effects in terms of a set of quantities, the tidal Love numbers~(TLNs), which encode the deformability properties of self-gravitating bodies~\cite{1909MNRAS..69..476L}. 
Initially exploited to study the structure of planets in the 
Solar System within Newtonian gravity, TLNs have been generalised to a fully relativistic description~\cite{Hinderer:2007mb,Binnington:2009bb,Damour:2009vw}, and have raised considerable attention in the context of 
binary neutron star~(NS) mergers, with the tantalising 
possibility of constraining the equation of state of dense 
matter from GW observations~\cite{Baiotti:2010xh,Baiotti:2011am,Vines:2011ud,Pannarale:2011pk,Vines:2010ca,Lackey:2011vz,Lackey:2013axa,Flanagan:2007ix,Favata:2013rwa,Yagi:2013baa,Maselli:2013mva,Maselli:2013rza,DelPozzo:2013ala,TheLIGOScientific:2017qsa,Bauswein:2017vtn, Most:2018hfd,Harry:2018hke,Annala:2017llu, Abbott:2018exr,Akcay:2018yyh,Abdelsalhin:2018reg, Jimenez-Forteza:2018buh, Banihashemi:2018xfb, Dietrich:2019kaq, Dietrich:2020eud, Henry:2020ski, Pacilio:2021jmq,Maselli:2020uol} (see
Refs.~\cite{GuerraChaves:2019foa,Chatziioannou:2020pqz} 
for some reviews). 

More recently, measurements of the TLNs from compact 
binaries have been proposed as a new tool to infer the 
properties of the environment in which the systems evolve. 
This possibility arises from the remarkable result that, within General Relativity, the TLNs of {\it naked} black holes~(BHs) (i.e., BHs in vacuum) vanish~\cite{Binnington:2009bb,Damour:2009vw,Damour:2009va,Pani:2015hfa,Pani:2015nua,Gurlebeck:2015xpa,Porto:2016zng,LeTiec:2020spy, Chia:2020yla,LeTiec:2020bos}. 
This property however is fragile, as it is broken for BH mimickers~\cite{Cardoso:2017cfl}, in extended theories of gravity~\cite{Cardoso:2017cfl,Cardoso:2018ptl,DeLuca:2022tkm}, in higher dimensions~\cite{Kol:2011vg,Cardoso:2019vof, Hui:2020xxx}, or in 
nonvacuum environments~\cite{Baumann:2018vus,Cardoso:2019upw,DeLuca:2021ite,Cardoso:2021wlq}.
In particular, during their cosmological evolution, environmental effects may provide an effective {\it dress} to BHs, resulting in nonzero TLNs. Such dresses may be 
formed around BHs due to secular effects like accretion\footnote{Accretion-driven dresses may be particularly important for primordial BHs, which are expected to be surrounded by a dark matter halo if they do not comprise the totality of the dark matter in the universe~\cite{Mack:2006gz,Adamek:2019gns}.
The detection of tidal effects can thus be used to distinguish primordial BHs (with their clouds) from other scenarios~\cite{Franciolini:2021xbq}.} or  
superradiant instabilities of ultralight bosonic fields~\cite{Brito:2015oca}, for which it has been shown 
that the TLNs would be proportional to inverse powers of the boson mass~\cite{Baumann:2018vus,DeLuca:2021ite}.

Love numbers of dressed BHs could be sufficiently large to 
leave an observable signature within GW signals emitted by coalescing binaries and, if measured, they could provide 
a smoking gun for the existence  of nonvacuum “structures” 
close to BHs. Unlike NS matter however, a low-density 
environmental dress may be unravelled during the last phases 
of the inspiral. In analogy with mass-shedding events occurring 
for stars with small compactness, disruption of the dress 
usually takes place when the binary semi-major axis is comparable with 
the Roche radius~\cite{Shapiro:1983du}. For smaller orbital radii, the environment progressively disappears and the 
coalescence proceeds with two naked BHs.
This process can be modelled through time-dependent TLNs, that smoothly 
approach zero.\footnote{Note that this is different from the notion of dynamical tidal deformability investigated in~\cite{Steinhoff:2016rfi,Creci:2021rkz,Consoli:2022eey} (see also~\cite{Nair:2022xfm}). In our case the TLNs are time-dependent because of the gradual disappearance of the environment around a BH, while the dynamical tidal deformability defined in~\cite{Steinhoff:2016rfi,Creci:2021rkz,Consoli:2022eey} measures the BH response to time-dependent tidal perturbations.
}
A similar phenomenon may occur when, during the 
common envelope phase of a neutron star binary, at least one 
of the objects with a mass very close to the Chandrasekhar 
limit turns into a BH~\cite{Singh:2022wvw}, changing its
TLNs from a finite value to zero.

The scope of this work is to study the relevance of such effect on BH binaries potentially 
observable by third-generation~(3G) detectors like the Einstein Telescope~(ET) or Cosmic Explorer~\cite{PhysRevD.91.082001,Hild:2010id,Sathyaprakash:2019yqt,Maggiore:2019uih,Essick:2017wyl,LIGOScientific:2016wof}, and to assess their ability to constrain the features of 
time-varying TLNs. We embed the latter within a BH waveform template properly 
built to model a smooth transition of the TLNs to zero, showing how few 
GW observations could shed new light on the matter content of the binary 
environment, and on the dynamical processes that lead to its tidal 
disruption. 
Hereafter we use geometric units, $G = c = 1$.

\section{Setup}
%------------------------------------------------------------------
We model the GW signal emitted by the binary using a modified version of the IMRPhenomD 
waveform model, which takes into account the inspiral, merger and 
ringdown phases of the coalescence~\cite{Husa:2015iqa,Khan:2015jqa}. 
The inspiral regime is further broken-up into an {\it early} and a 
{\it late} component, with the former being described by the 
post-Newtonian (PN) expansion, in which tidal effects add linearly 
to point-particle contributions and are fully encoded by the 
Love numbers. 
In Fourier space, the early part of the signal is given 
by~\cite{Sathyaprakash:1991mt,Damour:2000gg}, 
\be
\tilde h (f) = C_\Omega{\cal A}_\textnormal{PN}  e^{i \psi_\text{\tiny PP} (f) + i \psi_\text{\tiny Tidal} (f)},
\ee
where $\psi_\text{\tiny PP}$ contains terms up to the 3.5PN order~\cite{Damour:2000gg, Arun:2004hn, Buonanno:2009zt, Abdelsalhin:2018reg}, and depends on the binary chirp mass 
$\mathcal{M} = (m_1 m_2)^{3/5}/(m_1+m_2)^{1/5}$ and the 
symmetric mass ratio $\eta = \eta_1 \eta_2 = m_1 m_2/(m_1+m_2)^2$, 
where $m_{1,2}$ are the component masses. The phase 
$\psi_\text{\tiny PP}$ also includes linear spin terms up to 
3PN order through the (anti)symmetric combinations of 
the individual spin components $\chi_{s}=(\chi_{1}+\chi_{2})/2$ 
and $\chi_{a}=(\chi_{1}-\chi_{2})/2$, and quadratic spin 
corrections entering at 2PN order, which include the spin-induced 
quadrupole moments. For sake of simplicity we assume that 
the latter correspond to their (naked) Kerr values, and that the effect 
of the dress is subdominant.
An additional contribution to the GW phase comes from 
tidal heating, which depends on the energy absorbed at the 
horizon and is proportional to the BH cross section. This effect introduces a higher-order PN correction~\cite{Alvi:2001mx,Maselli:2017cmm}, 
which is typically small and negligible for our analysis, 
and hence we assume that is the same as for naked BHs~\cite{DeLuca:2021ite}.
The waveform amplitude ${\cal A}$, also 
expanded up to the 3PN order, depends on 
$({\cal M},\eta,\chi_{s},\chi_{a})$, with the 
leading term reading
\be
\mathcal{A}_\textnormal{PN} = \sqrt{\frac{5}{24}} 
\frac{\mathcal{M}^{5/6}f^{-7/6}}{\pi^{2/3}d_L} (1+\textnormal{PN corrections})\ ,
\ee
where $d_L$ is the luminosity distance. 
Finally the geometric factor $C_{\Omega}=[F_+^2 (1+\cos^2 \iota)^2 + 4 F_\times^2 \cos \iota]^{1/2}$ depends on the  
inclination $\iota$ which identifies the angle between the 
binary line of sight and its orbital angular momentum, 
and on the detector antenna pattern functions $F_{+,\times} 
(\theta, \varphi, \psi)$ which are 
functions  of the source position 
in the sky $(\theta, \varphi)$ and on the polarization angle $\psi$.

The dominant tidal correction enters the waveform at the
5PN order as $\psi_\text{\tiny Tidal} (f) = -\frac{39}{2}\tilde{\Lambda} (\pi M f)^{5/3}$, in terms of the total mass of the binary $M=m_1+m_2$ and of the effective tidal deformability parameter $\tilde \Lambda$, which depends on the masses and TLNs of each binary component~\cite{Flanagan:2007ix, Vines:2011ud}. We neglect higher-order PN contributions (starting at 6PN order) since they are hardly measurable and increase the dimensionality of the parameter space.

\begin{figure}[t!]
\centering
\includegraphics[width=0.44\textwidth]{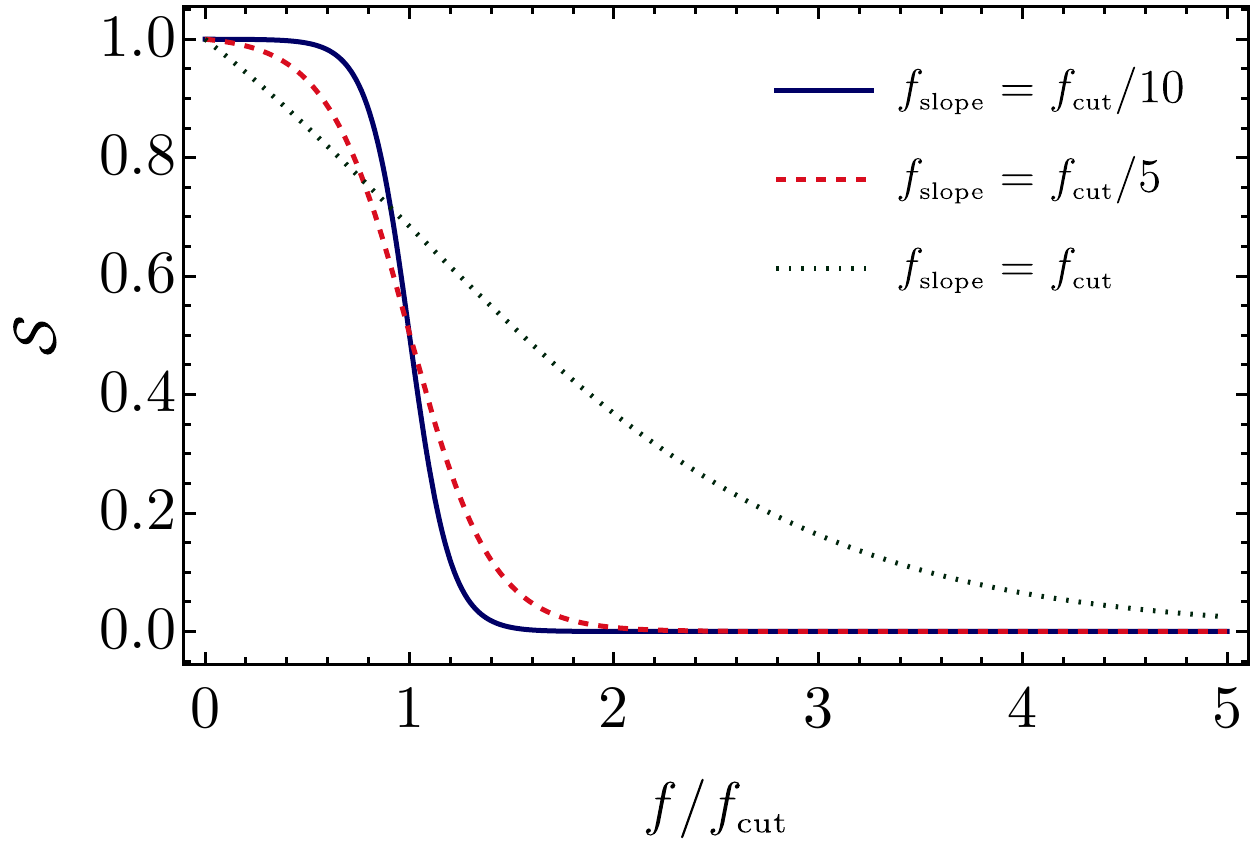}
\caption{\it Smoothing function 
${\cal S}(f)$ as a function of the frequency for different 
values of the slope parameter.}\label{smoothing}
\end{figure}

We assume that the halo surroundings the BHs is dynamically 
disrupted during the coalescence, i.e. that 
tidal effects disappear progressively 
from a characteristic cut-off frequency $f_\text{\tiny cut}$. 
To this aim we introduce a frequency-dependent tidal deformability 
\be
\tilde{\Lambda} \to {\cal S} (f) \cdot  \tilde{\Lambda} = \llp \frac{1+e^{-f_\text{\tiny cut}/f_\text{\tiny slope}}}{1+e^{(f-f_\text{\tiny cut})/f_\text{\tiny slope}}} \rrp \cdot \tilde{\Lambda}\ ,
\ee
which is cast in terms of a smoothing function ${\cal S}(f)$ 
that  approaches zero at frequencies larger than the cut-off 
with a characteristic slope $f_\text{\tiny slope}$. 
A representative example of the smoothing function is 
shown in Fig.~\ref{smoothing}. 
In our analysis tidal effects are therefore fully 
described by three quantities, $(\tilde \Lambda, f_\text{\tiny cut}, f_\text{\tiny slope})$, while the overall waveform model depends on 14 
parameters $\vec{\theta}=\{{\cal M},\eta,\chi_s,\chi_a,t_c,\phi_c,
d_L,\theta,\phi,\psi,\iota,\tilde \Lambda, f_\text{\tiny cut}, f_\text{\tiny slope}\}$, 
where $(t_c,\phi_c)$ are 
the time and phase at the coalescence.

\begin{figure*}[t!]
\centering
\includegraphics[width=1.02\textwidth]{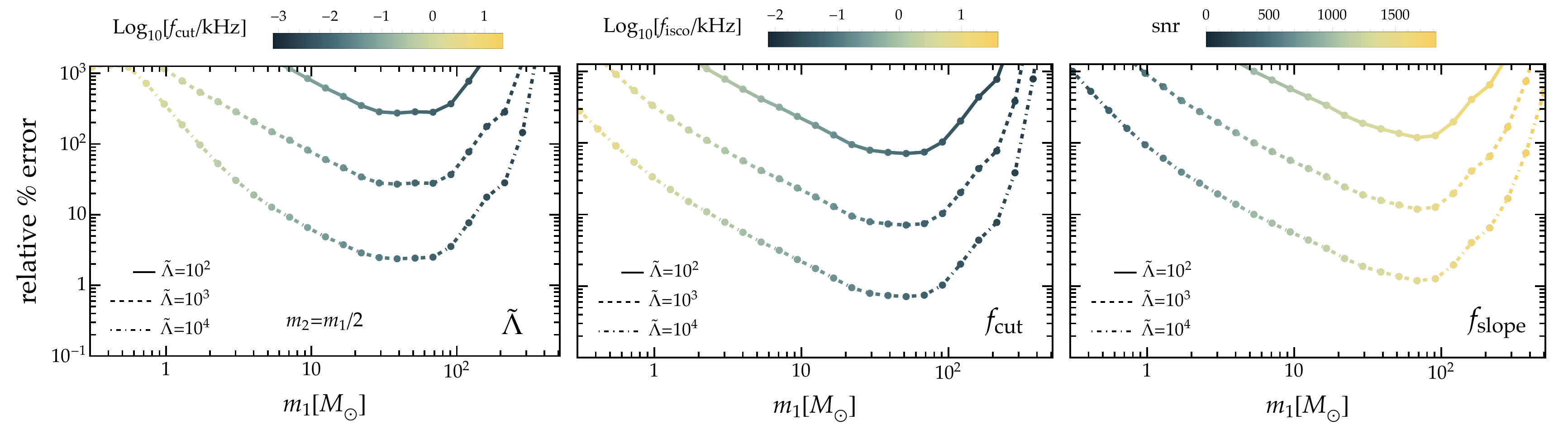}
\includegraphics[width=1.02\textwidth]{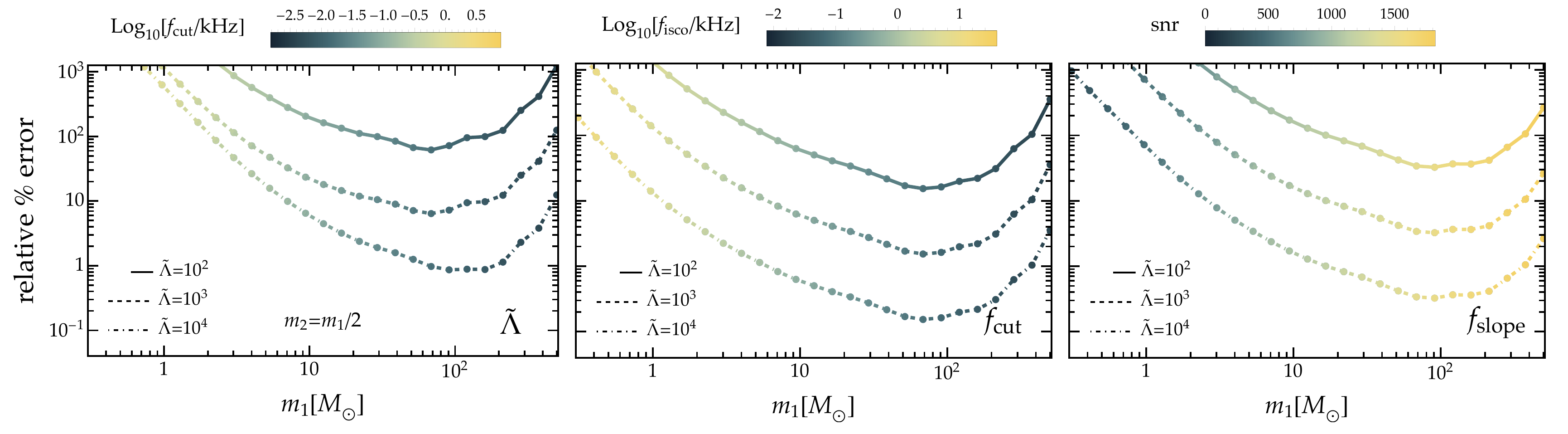}
\caption{\it (Top row) 
Relative percentage error on the effective tidal Love 
number (left), the cut-off frequency (center), and the 
slope frequency (right) for a binary with 
mass ratio $m_2 = m_1/2$, observed by ET at a 
luminosity distance of $d_L = 1 {\rm Gpc}$.
From left to right, the upper bars show the values of 
the cut-off frequency, of the ISCO frequency, and of the binary snr. Errors are computed 
assuming  $f_\text{\tiny cut} = f_\text{\tiny isco}/10$ 
and $f_\text{\tiny slope} = f_\text{\tiny cut}/5$.
(Bottom row) Same as top panels but choosing 
$f_\text{\tiny cut} = f_\text{\tiny isco}/5$ 
and $f_\text{\tiny slope} = f_\text{\tiny cut}/5$.}
\label{k2fcutfslope}
\end{figure*}

We study the detectability of the tidal parameters by using a 
Fisher-matrix approach~\cite{Poisson:1995ef, Vallisneri:2007ev, Cardoso:2017cfl}. 
For signals with large signal-to-noise ratio~(snr), as those expected 
for 3G detectors, the posterior distribution of $\vec{\theta}$ 
can be described by a multivariate Gaussian distribution centered 
around the {\it true} values $\vec{\hat{\theta}}$, 
with covariance ${\bf \Sigma} = {\bf \Gamma}^{-1}$, where
\be
\Gamma_{ij}= \left\langle \frac{\partial h}{\partial \theta_i}\bigg\vert\frac{\partial h}{\partial \theta_j}\right\rangle_{\vec{\theta}=\vec{\hat{\theta}}}
\ee
is the Fisher information matrix, 
and  statistical error on the $i$-th parameter is 
given by $\sigma_i=\Sigma^{1/2}_{ii}$.
In the previous expression we have introduced the scalar 
product over the detector noise spectral density $S_n(f)$ 
between two waveform templates $h_{1,2}$:
\be
\langle h_1\vert 
h_2\rangle=4\Re\int_{f_\text{\tiny min}}^{f_\text{\tiny max}} \frac{\tilde{h}_1(f)\tilde{h}^\star_2(f)}{S_n(f)}df \ ,\label{scalprod}
\ee
where $\star$ denotes complex conjugation. The snr of a given 
signal is ${\rm snr}=\langle h\vert h\rangle^{1/2}$.
In our analysis we fix the minimum and maximum frequency of 
integration to $f_\text{\tiny min}=1$Hz and 
$f_\text{\tiny max}=1.2f_\text{\tiny RD}$, where $f_\text{\tiny RD}$ is the remnant's ringdown 
frequency~\cite{Khan:2015jqa}.

Hereafter we consider optimally-oriented binaries, 
removing the four angles from the Fisher analysis, thus 
reducing ${\bf \Gamma}$ to a $10\times 10$ square matrix.
We also fix $\chi_1=\chi_2=0$ and $t_c=\phi_c=0$. We assume 
that dressed binary BHs are observed by ET adopting the design ET-D sensitivity curve~\cite{Hild:2010id}.

\begin{figure*}[t!]
\centering
\includegraphics[width=0.5\textwidth]{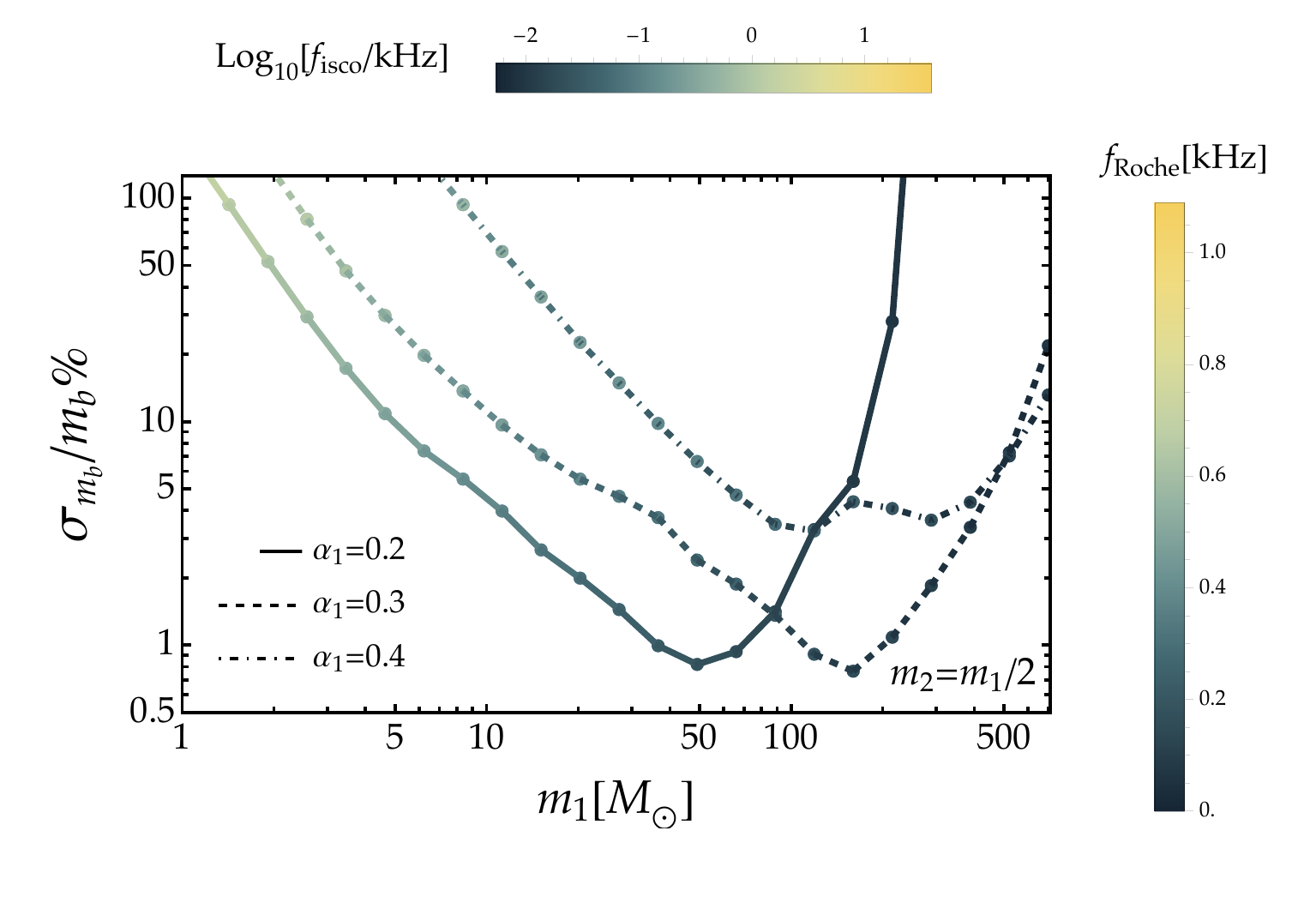}
\includegraphics[width=0.43\textwidth]{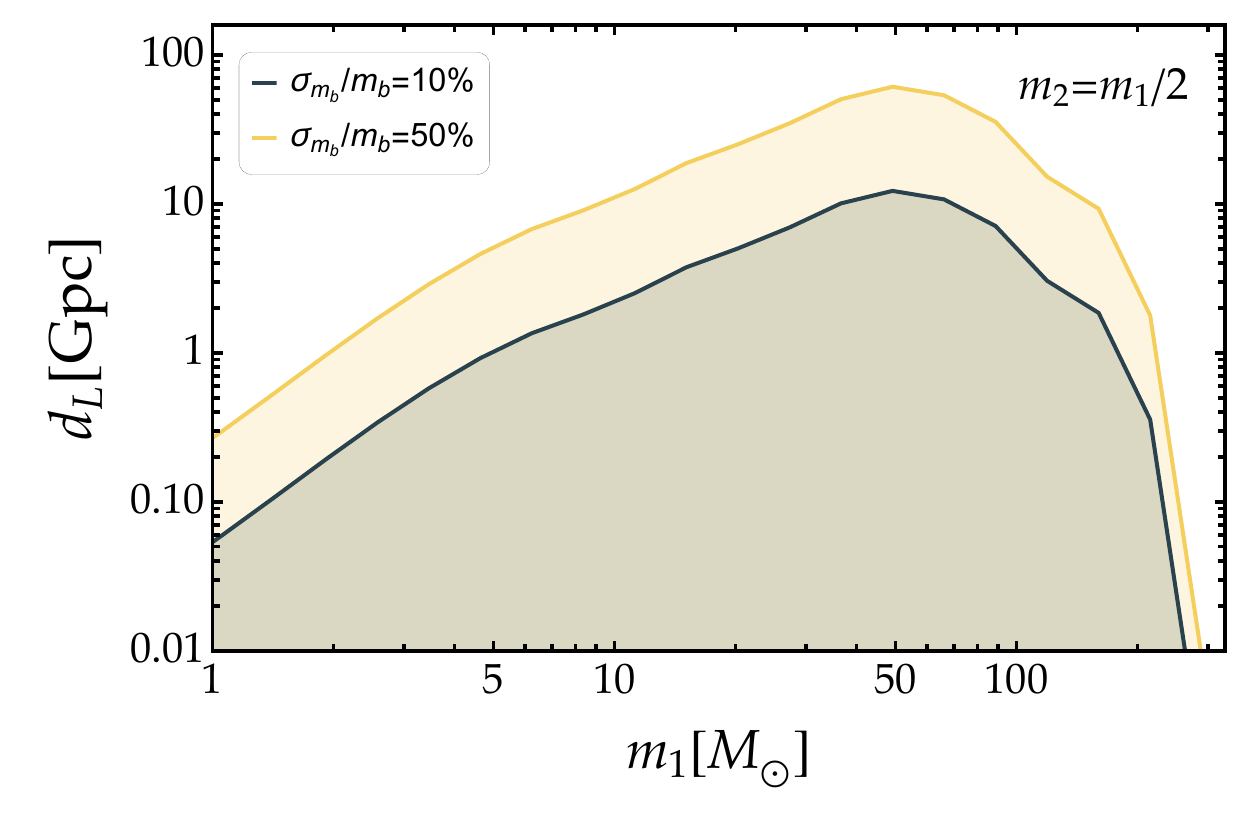}
\caption{\it (Left): Relative percentage error on the 
scalar field mass $m_b$ for a
binary system with primary mass $m_1$ and mass ratio $m_2 = m_1/2$, 
detected by ET at a luminosity distance of 
$d_L = 1 \, {\rm Gpc}$. Upper and lateral bars 
show the values of the ISCO and of the Roche frequency, 
respectively. Errors are computed assuming 
$f_\text{\tiny slope} = f_\text{\tiny Roche}/5$. 
(Right) Maximum luminosity distance for which the 
scalar field mass $m_b$, assuming $\alpha_1 = m_1  m_b =0.2$, can be constrained by ET with a relative percentage accuracy of 
$10\%$ (green) and $50\%$ (yellow).}
\label{alpha}
\end{figure*}

%------------------------------------------------------------------
% \noindent{{\bf{\em Results.}}}
\section{Results}
%------------------------------------------------------------------

\subsection{General framework}
We apply the framework discussed above to investigate the 
detectability of the tidal parameters. The results of the Fisher 
analysis are shown in Fig.~\ref{k2fcutfslope}, 
for different choices of $\tilde \Lambda =(10^2,10^3,10^4)$. 
Larger values would break the convergence of the post-Newtonian 
expansion, leading the tidal term to dominate over the lower 
PN orders. For simplicity we focus on binaries with mass ratio
$m_2/m_1=1/2$, although both masses are included as waveform parameters of the Fisher analysis. Different choices provide similar results.
The ISCO frequency $f_\text{\tiny isco}$ for each binary is computed taking into account self-force effects due to the lighter component of the binary~\cite{Favata:2010ic}.

The top panels of Fig.~\ref{k2fcutfslope} assume $f_\text{\tiny cut} = f_\text{\tiny isco}/10$ 
and $f_\text{\tiny slope} = f_\text{\tiny cut}/5$ for the cut-off 
and slope frequency, respectively. 
Note that for all binary configurations analysed we have checked that 
the cut-off frequency is (well) below the frequency describing the 
transition between the early and the late inspiral, 
$f_\text{\tiny ei-li} \simeq 0.018/M_\BH$~\cite{Husa:2015iqa,Khan:2015jqa}.
The panels of Fig.~\ref{k2fcutfslope} show that all the 
tidal parameters can be measured with high accuracy for sufficiently large values of the effective tidal Love number.
In particular, considering a dressed 
BH system with primary mass $m_1 = 50 M_\odot$ and mass ratio $m_2 = m_1/2$ at a distance $d_L = 1 \,{\rm Gpc}$, 
ET will be able to measure its tidal deformability with a relative 
accuracy of a few percent and an snr of a few hundred. 
A similar precision can be reached for the cut-off and 
the slope frequency of the smoothing function, showing the 
potential of ET in measuring the transition point where the 
TLN vanishes. These results slightly improve if we assume 
$f_\text{\tiny cut} = f_\text{\tiny isco}/5$ and 
$f_\text{\tiny slope} = f_\text{\tiny cut}/5$, as shown in the 
bottom panels of Fig.~\ref{k2fcutfslope}. This is expected 
since larger value of the cut-off frequency translates into a 
longer inspiral phase in which tidal effects are active.
Changing the slope frequency does not significantly modify the 
results of our analysis. For example, we have checked that a 
decrease in $f_\text{\tiny slope}$, which results into a sharper cut-off, 
does not strongly impact on the estimated errors.

\subsection{First-principle model: BHs dressed by ultralight bosons}
We can now focus on a specific example of a  
dress sourced by an ultralight bosonic scalar field with mass $m_b$ undertaking a phase of accretion or superradiant instability~\cite{Brito:2015oca} onto a BH. 
In this case it was shown that the TLN of each body is proportional 
to the inverse of the gravitational coupling 
$\alpha_i = m_i m_b$, namely~\cite{DeLuca:2021ite}
\begin{align}
\tilde{\Lambda} = \frac{32}{39} \lp \frac{12}{\eta_1} - 11 \rp \eta_1^5 k_2^{(1)} + (1 \leftrightarrow 2)\ ,
\end{align}
where $k_2^{(i)} \propto 1/\alpha_i^8$.
For this model, the effective tidal deformability vanishes when the clouds surrounding each binary component are tidally disrupted, that is when the binary semi-major axis is comparable to the Roche radius. For different binary components this results into different cut-off frequencies $f_\text{\tiny cut}^i$ for each body. We therefore introduce an effective frequency-dependent TLN as $k_2^{(i)} \to S^i(f) k_2^{(i)}$, where each smoothing function $S^i(f)$ is determined by the same slope frequency  $f_\text{\tiny slope}$, but by different cut-off frequencies~\cite{DeLuca:2021ite}
\be
\label{roche}
f_\text{\tiny Roche} (\alpha_i, m_i) = \frac{3 \sqrt{3}}{\pi \gamma^{3/2}} \alpha_i^3 \lp \frac{M_s}{m_i} \rp^{1/2} f_\text{\tiny isco}\ ,
\ee
in terms of a numerical coefficient $\gamma \sim \mathcal{O}(2)$ 
which takes values from 1.26 for rigid bodies to 2.44 for 
fluid ones, and of the total mass enclosed in the scalar 
cloud $M_s$. We fix the latter to the upper bound 
$M_s = 0.1 m_i$, which saturates the regime of validity 
of the perturbative expansion. Note that, in a first-principle model such as this one, the number of waveform parameters to be constrained is smaller than in the general model, since the gravitational couplings $\alpha_i$ dictate
both the TLNs and the cut-off frequencies.

The results for the Fisher analysis for this case are
shown in Fig.~\ref{alpha}. The left panel shows the 
relative percentage error on the gravitational coupling 
as a function of the primary BH mass, assuming a luminosity distance 
of $d_L = 1 \, {\rm Gpc}$ and mass ratio $m_2/m_1 = 1/2$. The upper and lateral bars 
identify the corresponding ISCO and Roche frequencies, respectively. One can appreciate that in the mass range 
$m_1 \in (2 \divisionsymbol 200) M_\odot$ ET could 
measure the coupling with an accuracy of a few percent.
This result improves the one discussed 
in~\cite{DeLuca:2021ite} where the 
integration in Eq.~\eqref{scalprod} was cut at the 
ISCO frequency, also assuming the signal amplitude to 
vanish afterward. 
Frequencies after the cut-off, which can be consistently  
taken into account within the PhenomD waveform 
model, provide here a significant boost to the 
parameter's reconstruction, which in turn translates 
into an improvement on the errors of the 
tidal parameters.\footnote{Note that, within our general model, the results of~\cite{DeLuca:2021ite} can be recovered in the $f_\text{\tiny slope}/f_\text{\tiny cut}\ll1$ limit, where effectively the tidal part of the signal vanishes for $f>f_\text{\tiny cut}$.}
In the right panel of Fig.~\ref{alpha} we show the maximum luminosity distance for which ET would be able to constrain the scalar field 
mass $m_b$ with a relative percentage error of $10\%$ (green line) and 
$50\%$ (yellow line), assuming a gravitational coupling $\alpha_1 = 0.2$. 
The result shows that ET could measure the scalar field mass 
with high accuracy within distances of  a
few Gpc.

Finally, let us stress that, as tidal interactions take place and lead to the disruption of the clouds, the effective mass of the BH+halo system decreases, probably resulting into a naked binary system coalescing within an unbound and more diluted halo.
As a proof of concept, in order to estimate the effect of this mass decrease in the waveforms, we have introduced a similar suppression factor $S(f)$ as the one discussed above to the BH masses, i.e. $m_i \to S(f) \, m_i$, which interpolates between unity and $0.9$, as the cloud's mass accounts at most for about 10\% of the BH masses in relevant scenario like superradiance. By introducing this frequency-dependent factor for the masses through all PN orders, we found that the error on $\tilde{\Lambda}$ increases by only a factor of 2, while the errors on $f_\text{\tiny cut}$ and $f_\text{\tiny slope}$ become much smaller, as these parameters appear also at lower PN orders and are therefore better constrained. 
Another interesting extension is to account for the orbital energy loss associated with the halo disruption~\cite{Baumann:2021fkf,Baumann:2022pkl}.
We leave a better investigation of these interesting effects to future work.

%----------------------------------------------------------------------------------------------------
% \noindent{{\bf{\em Conclusions.}}}
\section{Conclusions}
%----------------------------------------------------------------------------------------------------
% \noindent
Tidal interactions active during the inspiral phase 
of binary coalescences carry unique information on the 
internal structure of compact objects {\it and} 
on the properties of the surroundings in which the 
systems evolve.
BHs dressed by a halo formed by dark matter, by a phase of accretion, or by the superradiant instability of
a ultralight boson, provide some astrophysical scenarios 
in which the external environments leave a footprint  
on the emitted GW signal~\cite{Cole:2022fir}, encoded in particular in nonvanishing TLNs~\cite{Baumann:2018vus,Cardoso:2019upw,DeLuca:2021ite,Cardoso:2021wlq}. In these cases, however, 
the gravitational interaction with the BH companion 
may destroy the halo, leading to tidal effects that fade away in the waveform at higher frequency. 

Motivated by such examples, in this paper we have 
investigated the potential of 3G detectors like ET 
to measure the variation of the Love number once the 
binary inspiral crosses a given cut-off frequency. 
To this aim we have augmented the standard IMRphenomD 
waveform with the inclusion of an effective tidal 
deformability which interpolates the transition between 
a finite effective Love number $\tilde \Lambda$ and the full vacuum regime.
We have carried out a statistical analysis on a wide 
range of sources observable by ET, finding that, together 
with $\tilde \Lambda$, both the cut-off frequency and the slope of 
the effective deformability are potentially measurable, 
with an accuracy of a few percent for stellar mass binaries.
The simultaneous measurements of all tidal parameters 
would provide key information on the composition of the 
BH environment, and on the mechanisms responsible 
for the vanishing of the Love number.

These results also pave the avenue for multiband analyses 
between ET and LISA~\cite{LISA:2017pwj}, for sources in a specific 
mass range~\cite{2016PhRvL.116w1102S, Vitale:2016rfr}.
For example, one can envisage a scenario in which a 
binary system of two dressed BHs with 
masses $m_1 \sim m_2 \sim 10^2 M_\odot$ inspirals within the 
frequency band of LISA, with a cut-off frequency 
falling into the ET bandwidth. 
A statistical analysis for LISA only would suggest that 
all the parameters of the effective tidal 
deformability are unmeasurable. However, a joint 
analysis which takes into account that the binary keeps 
evolving in ET with fading Love number, would 
dramatically improve the parameter reconstruction. 
For a $(10^2$-$10^2)M_\odot$ binary, and assuming 
$\tilde \Lambda = 10^3$, the multiband analysis would lead 
to a stringent constraint on the 
environment Love number at the level of $\sigma_{\tilde \Lambda}/{\tilde \Lambda} \approx 1\%$. 

Finally, let us stress that a frequency-dependent TLN can be used to disentangle dressed BHs from neutron stars in the solar mass range, since the latter are characterised by tidal effects which persist until the last stages of the merger. We plan to investigate this point in a future work.

%----------------------------------------------------------------------------------------------------
% \noindent{{\bf{\em Acknowledgments.}}}
%----------------------------------------------------------------------------------------------------
\begin{acknowledgments}
We thank G.~Franciolini for interesting comments.
A.M. thanks the University of Pennsylvania for the hospitality during the completion of this project.
V.DL. is supported by funds provided by the Center for Particle Cosmology at the University of Pennsylvania. 
A.M. acknowledges financial support from the 
EU Horizon 2020 Research and Innovation 
Programme under the Marie Sklodowska-Curie Grant 
Agreement no. 101007855.
P.P. acknowledges support provided under the European Union’s H2020 ERC, Starting Grant agreement no. DarkGRA–757480 and under the MIUR PRIN programme, and support from the Amaldi Research Center funded by the MIUR program ``Dipartimento di Eccellenza'' (CUP: B81I18001170001).
\end{acknowledgments}

\bibliography{draft}

\end{document}